\documentclass[pra,aps,amsmath,amssymb,showpacs,showkeys,twocolumn,nofootinbib]{revtex4}
\newcommand{\bsg}{\boldsymbol{\sigma}}
\newcommand{\bro}{\boldsymbol{\rho}}
\newcommand{\vro}{\boldsymbol{\varrho}}
\newcommand{\bom}{\boldsymbol{\omega}}
\newcommand{\pen}{\openone}
\newcommand{\hw}{\widetilde{H}}

\newcommand{\rh}{{\textup{H}}}
\newcommand{\Tr}{{\textup{Tr}}}
\newcommand{\spc}{{\textup{spec}}}
\newcommand{\az}{\boldsymbol{Z}}
\newcommand{\ax}{\boldsymbol{X}}
\newcommand{\ppsf}{\boldsymbol{P}}
\newcommand{\qqsf}{\boldsymbol{Q}}
\newcommand{\cle}{\mathcal{E}}

\begin{document}
\clearpage
\preprint{}

\title{Uncertainty and certainty relations for successive projective measurements of a qubit in terms of Tsallis' entropies}
\author{Alexey E. Rastegin}
\email{rast@api.isu.ru; alexrastegin@mail.ru}
\affiliation{Department of Theoretical Physics, Irkutsk State University,
Gagarin Bv. 20, Irkutsk 664003, Russia}

\begin{abstract}
We study uncertainty and certainty relations for two successive
measurements of two-dimensional observables. Uncertainties in
successive measurement are considered within the following two
scenarios. In the first scenario, the second measurement is
performed on the quantum state generated after the first
measurement with completely erased information. In the second
scenario, the second measurement is performed on the
post-first-measurement state conditioned on the actual measurement
outcome. Induced quantum uncertainties are characterized by means
of the Tsallis entropies. For two successive projective
measurement of a qubit, we obtain minimal and maximal values of
related entropic measures of induced uncertainties. Some
conclusions found in the second scenario are extended to arbitrary
finite dimensionality. In particular, a connection with mutual
unbiasedness is emphasized.
\end{abstract}


\maketitle

\section{Introduction}\label{sec1}

The Heisenberg uncertainty principle \cite{wh27} is one of the
fundamentals of quantum theory. Despite of its wide popularity,
there is no general consensus over the scope and
validity \cite{lahti,hall99}. It is typically said that measuring
some observable will inevitably disturb the system, whence the
context for further observations is raised. There are many ways
to quantify uncertainties as well as few scenarios of measuring
observables. The first form of explicit mathematical formulation
has been given for the position and momentum by Kennard
\cite{kennard}. For this pair, a product of the
standard deviations cannot be less than $\hbar/2$. For any pair of
observables, this direction was realized by Robertson
\cite{robert}. However, the traditional approach deals with
quantum uncertainties raised in two different experiments with the
same pre-measurement state. So, this approach does not reveal many
details about a disturbance of the system due to performed
measurements. Rather, Heisenberg's initial reasons are better
formulated in terms of noise and disturbance \cite{ozawa04}. As
was discussed in Refs. \cite{mdsrin03,paban13}, studies of quantum
uncertainties in the results of successive measurements have
received less attention than they deserve.

Much attention to uncertainty relations is stimulated by a
progress in using quantum systems as informational recourses
\cite{ww10}. Hence, formulations in information-theoretic terms
including entropies are of interest. Traditional uncertainty
relations for quantum measurements deal with probability
distributions calculated for one and the same input state. This
treatment was studied in entropic terms
\cite{deutsch,maass,brud11} and recently by means of majorization
technique \cite{prz13,fgg13,rpz14}. In quantum information
processing, other situations are rather typical. Our subsequent
manipulations deal with an output state of the latter stage. In
the case of two successive measurements, the following two
scenarios are typically addressed \cite{bfs2014,zzhang14}. In the
first scenario, the second measurement is performed on the quantum
state generated after the first measurement with completely erased
information. In the second scenario, the second measurement is
performed on the post-first-measurement state conditioned on the
actual measurement outcome. Thus, the scenarios are related to a
realistic situation, when subsequent actions deal with an output
state of the latter stage.

In the present work, we study uncertainties in successive
measurements with the use of Tsallis entropies. We also focus on
maximal possible values of related entropic measures. Indeed,
certainty relations for successive measurements seem to be not
considered previously. For successive measurements of a pair of
qubit observables, we obtain uncertainty and certainty entropic
bounds. The paper is organized as follows. In Sec. \ref{sec2}, the
background material is reviewed. Here, definitions of the used
entropic functions are recalled. In Sec. \ref{sec3}, we generally
discuss quantum uncertainties induced by successive measurements
of a pair of observables. Two scenarios of such measurements and
related $\alpha$-entropic measures are introduced. In Section
\ref{sec4}, we derive tight uncertainty and certainty relations
for a qubit within the first scenario. The conditions for equality
are obtained and discussed. Tight lower and upper bounds on the
conditional $\alpha$-entropies within the second scenario are
presented in Sec. \ref{sec5}. We also observe a connection of the
equality conditions for upper bounds with mutual unbiasedness.
This observation remains valid in arbitrary finite dimensions. In
Sec. \ref{sec6}, we conclude the paper with a summary of results.

\section{Preliminaries}\label{sec2}

To quantify uncertainties in generated probability distributions,
we will use Tsallis' entropies. Let discrete random variable $X$
take values on a finite set $\Omega_{X}$ of cardinality
$\#\Omega_{X}$. The Tsallis entropy of degree $\alpha>0\neq1$ is
defined by \cite{tsallis}
\begin{equation}
H_{\alpha}(X):=\frac{1}{1-\alpha}\left(\sum_{{\,}x\in\Omega_{X}} p(x)^{\alpha}
- 1\right)
{\,}. \label{tsaent}
\end{equation}
With other factor, this function was examined by Havrda and
Charv\'{a}t \cite{havrda} and later by Dar\'{o}czy \cite{ZD70}. In
statistical physics, the entropy (\ref{tsaent}) is extensively
used due to Tsallis \cite{tsallis}. For other multidisciplinary
applications, see the book \cite{gmt} and references therein. The
R\'{e}nyi entropies \cite{renyi61} form another especially
important family of one-parametric extensions of the Shannon
entropy. For $\alpha>0\neq1$, R\'{e}nyi's $\alpha$-entropy can be
expressed via Tsallis' entropy as
\begin{equation}
R_{\alpha}(X)=\frac{1}{1-\alpha}{\>}
\ln\bigl(1+(1-\alpha)H_{\alpha}(X)\bigr)
\ . \label{exrt}
\end{equation}
Despite of a direct relation, the entropies (\ref{tsaent}) and
(\ref{exrt}) differ in essential properties. As a rule,
formulation in terms of one of the two entropies cannot
immediately be recast for other. Such a situation takes place for
entropic uncertainty relations too. In Ref. \cite{rast14a}, uncertainty 
and certainty relations for the Pauli observables were derived
in terms of the R\'{e}nyi entropies. Using R\'{e}nyi's entropies,
uncertainty relations for successive projective measurements were 
studied in Ref. \cite{zzhang14}. Properties and applications of both 
the types of entropies in quantum theory are considered in the book 
\cite{bengtsson}.

Obviously, the function $\bigl(\xi^{\alpha}-\xi\bigr)/(1-\alpha)$
is concave for all $\alpha>0$. Hence, the entropy (\ref{tsaent})
is a concave function of probability distribution. The maximal
value $\ln_{\alpha}(\#\Omega_{X})$ is reached by the uniform
distribution. It is often convenient to rewrite Eq. (\ref{tsaent})
as
\begin{align}
H_{\alpha}(X)&=-\sum_{x\in\Omega_{X}}p(x)^{\alpha}{\,}\ln_{\alpha}{p}(x)
\nonumber\\
&=\sum_{x\in\Omega_{X}}p(x){\>\,}{\ln_{\alpha}}{\left(\frac{1}{p(x)}\right)}
\, . \label{tsaln}
\end{align}
Here, we used the $\alpha$-logarithm defined for $\alpha>0\not=1$
and $\xi>0$ as
\begin{equation}
\ln_{\alpha}(\xi)=\frac{\xi^{1-\alpha}-1}{1-\alpha}
\ . \label{lnadf}
\end{equation}
In the limit $\alpha\to1$, we have $\ln_{\alpha}(\xi)\to\ln\xi$,
so that the Shannon entropy $H_{1}(X)=-\sum_{x}p(x)\ln{p}(x)$ is
raised. For brevity, we will typically write entropic sums
without set symbols such as $\Omega_{X}$.

The above definitions are also applied in the quantum regime. Let
the state of a quantum system be described by the density matrix
$\bro$. It is a positive semi-definite matrix with the unit trace.
The quantum $\alpha$-entropy of $\bro$ is defined as
\begin{equation}
\rh_{\alpha}(\bro):=\frac{1}{1-\alpha}{\>}\Bigl(\Tr(\bro^{\alpha})-1\Bigr)
\, . \label{qtsdf}
\end{equation}
In the case $\alpha=1$, we deal with the von Neumann entropy
$\rh_{1}(\bro)=-\Tr(\bro\ln\bro)$. For general properties on the
von Neumann entropy, see Refs. \cite{wehrl,ohya}.
In the following, the entropies (\ref{tsaent}) and (\ref{qtsdf})
will be used in studying uncertainties in successive projective
measurements.

In one of the two scenarios of successive measurements, the
second measurement is performed on the actual
post-first-measurement state. Analyzing this
scenario, conditional form of entropies will be utilized. The
conditional entropies are widely used in information theory
\cite{CT91} and in applied disciplines. The standard conditional
entropy is defined as follows. Let $X$ and $Z$ be random
variables. For each $z\in\Omega_{Z}$, we take the function
\begin{equation}
H_{1}(X|z)=-\sum_{x\in\Omega_{X}}p(x|z){\,}\ln{p}(x|z)
\ . \label{csheny}
\end{equation}
By $p(x|z)$, we mean the conditional probability that $X=x$ given
that $Z=z$. By Bayes' rule, it obeys $p(x|z)=p(x,z)/p(z)$. The
entropy of $X$ conditional on knowing $Z$ is defined as
\cite{CT91}
\begin{align}
H_{1}(X|Z)&:=\sum_{z\in\Omega_{Z}} p(z){\,}H_{1}(X|z)
\nonumber\\
&=-\sum_{x\in\Omega_{X}}\sum_{z\in\Omega_{Z}} p(x,z){\,}\ln{p}(x|z)
\ . \label{cshen}
\end{align}
For brevity, we will further write entropic sums without
mentioning that $x\in\Omega_{X}$ and $z\in\Omega_{Z}$. In the
context of quantum theory, the conditional entropy (\ref{cshen})
was used in information-theoretic formulations of Bell's theorem
\cite{BC88} and noise-disturbance uncertainty relations
\cite{bhow13}.

In the literature, two kinds of the conditional THC entropy were
considered \cite{sf06,rastkyb}. These forms are respectively
inspired by the two expressions shown in Eq. (\ref{tsaln}). The
first conditional form is defined as \cite{sf06}
\begin{equation}
H_{\alpha}(X|Z):=\sum_{z} p(z)^{\alpha}{\,}H_{\alpha}(X|z)
\ , \label{hct1}
\end{equation}
where
\begin{equation}
H_{\alpha}(X|z):=\frac{1}{1-\alpha}{\,}\left(\sum_{x} p(x|z)^{\alpha}-1\right)
{\,}. \label{hctm0}
\end{equation}
The conditional entropy (\ref{hct1}) is, up to a factor, the
quantity originally introduced by Dar\'{o}czy \cite{ZD70}. For any
$\alpha>0$, the conditional entropy (\ref{hct1}) satisfies the
chain rule \cite{sf06,ZD70}. The chain rule is essential in some
applications of conditional entropies in quantum information
science. For example, this rule for the standard conditional
entropy is used in deriving the Braunstein--Caves inequality
\cite{BC88}. This inequality expresses an entropic version of the
Bell theorem \cite{bell64}. The $\alpha$-entropies were applied in
formulating non-locality, contextuality and non-macrorealism
inequalities \cite{rastqic14,rastq14}.

Using the particular function (\ref{hctm0}), the second form of
conditional $\alpha$-entropy is written as \cite{sf06,rastkyb}
\begin{equation}
\hw_{\alpha}(X|Z):=\sum_{z} p(z){\,}H_{\alpha}(X|z)
\ . \label{hct2}
\end{equation}
It should be noted that this form of conditional entropy does not
share the chain rule of usual kind \cite{sf06}. Nevertheless, the
conditional entropy (\ref{hct2}) is interesting at least as an
auxiliary quantity \cite{rastkyb}. Moreover, it can be used in
studying the Bell theorem \cite{rastrian}, even though the chain
rule is not applicable here.

\section{On successive projective measurements in finite dimensions}\label{sec3}

In this section, we consider several facts concerning uncertainty
and certainty relations for successive measurements in finite
dimensions. Let $\az$ be an observable of some finite-level
quantum system. By $\spc(\az)$, we denote the spectrum of $\az$.
The spectral decomposition is written as
\begin{equation}
\az=\sum_{z\in\spc(\az)}{z{\,}\ppsf_{z}}
\ , \label{xspc}
\end{equation}
where $\ppsf_{z}$ denotes the orthogonal projection on the
corresponding eigenspace of $\az$. The operators $\ppsf_{z}$ are
mutually orthogonal and satisfy the completeness relation
\begin{equation}
\sum_{z\in\spc(\az)}{\ppsf_{z}}=\pen
\ . \label{cmrl}
\end{equation}
By $\pen$, we denote the identity operator in the Hilbert space of
studied system. Let the pre-measurement state be described by the
density matrix $\bro$. The probability of each outcome $z$ is
expressed as $\Tr(\ppsf_{z}{\,}\bro)$. Due to Eq. (\ref{cmrl}),
these probabilities are summarized to $1$. Calculating the
$\alpha$-entropy of the generated probability distribution, we
will deal with the quantity
\begin{equation}
H_{\alpha}(\az;\bro)=\frac{1}{1-\alpha}
\left(\sum_{{\,}z\in\spc(\az)} \bigl(\Tr(\ppsf_{z}{\,}\bro)\bigr)^{\alpha}
- 1\right)
{\,}.   \label{haby}
\end{equation}
Let $\ax$ be another observable with the spectral decomposition
\begin{equation}
\ax=\sum_{x\in\spc(\ax)}{x{\,}\qqsf_{x}}
\ . \label{yspc}
\end{equation}
Here, the operator $\qqsf_{x}$ is the orthogonal projection on the
corresponding eigenspace of $\ax$. We consider an amount of
uncertainties induced by two successive measurements of the
observables, $\az$ at first and $\ax$ later. As was above
mentioned, there are two possible scenarios of interest.

In the first scenario, the second measurement is performed on the
quantum state generated after the first measurement with
completely erased information. That is, the second measurement is
performed with the pre-measurement state
\begin{equation}
\cle_{\az}(\bro)=\sum_{{\,}z\in\spc(\az)}\ppsf_{z}{\,}\bro{\,}\ppsf_{z}
\ . \label{smpm}
\end{equation}
The linear map $\cle_{\az}$ describes the action of the projective
measurement of $\az$. The right-hand side of Eq. (\ref{smpm}) is
an operator-sum representation of the map \cite{nielsen}. The
completeness relation (\ref{cmrl}) provides that the map
(\ref{smpm}) preserves the trace for all inputs. The uncertainty
in the second measurement is quantified by means of the entropy
$H_{\alpha}\bigl(\ax;\cle_{\az}(\bro)\bigr)$. The latter is
expressed similarly to Eq. (\ref{haby}), but with the
probabilities $\Tr\bigl(\qqsf_{x}{\,}\cle_{\az}(\bro)\bigr)$. We
will characterize a total amount of uncertainty in the first
scenario by the sum of the classical entropies of two generated
probability distributions. Note that the post-first-measurement
state (\ref{smpm}) obeys the following property. If we have
measured $\az$ in the state $\cle_{\az}(\bro)$, we again deal with
probabilities $\Tr(\ppsf_{z}{\,}\bro)$. Thus, we write the formula
\begin{align}
&H_{\alpha}(\az;\bro)+H_{\alpha}\bigl(\ax;\cle_{\az}(\bro)\bigr)
\nonumber\\
&=\rh_{\alpha}\bigl(\cle_{\az}(\bro)\bigr)+\rh_{\alpha}\bigl(\cle_{\ax}\circ\cle_{\az}(\bro)\bigr)
\ . \label{hhcq}
\end{align}
By $\cle_{\ax}\circ\cle_{\az}$, we mean the composition of two
quantum operations \cite{nielsen}. Further, the projectors
$\qqsf_{x}$ give rise to the map
\begin{equation}
\cle_{\ax}(\bom)=\sum_{{\,}x\in\spc(\ax)}\qqsf_{x}{\,}\bom{\,}\qqsf_{x}
\ , \label{pmsm}
\end{equation}
where $\bom$ is a density matrix. The left-hand side of Eq.
(\ref{hhcq}) is the sum of classical $\alpha$-entropies of the two
probability distributions, whereas the right-hand side is the sum
of quantum $\alpha$-entropies. To formulate uncertainty and
certainty relations for successive measurements, we aim to have a
two-sided estimate on the quantity (\ref{hhcq}).

Let us recall one of physically important properties of the von
Neumann entropy related to the measurement process. In effect,
projective measurements cannot decrease the von Neumann entropy
(see, e.g., theorem 11.9 in Ref. \cite{nielsen}), that is
\begin{equation}
\rh_{1}\bigl(\cle_{\az}(\bro)\bigr)\geq\rh_{1}(\bro)
\ . \label{vnpm}
\end{equation}
In the paper \cite{rastjst}, we extended the above property to the
family of quantum unified entropies. In particular, for all
$\alpha>0$ we have \cite{rastjst}
\begin{equation}
\rh_{\alpha}\bigl(\cle_{\az}(\bro)\bigr)\geq\rh_{\alpha}(\bro)
\ . \label{tepm}
\end{equation}
It follows from Eqs. (\ref{hhcq}) and (\ref{tepm}) that
\begin{equation}
H_{\alpha}(\az;\bro)+H_{\alpha}\bigl(\ax;\cle_{\az}(\bro)\bigr)\geq
\rh_{\alpha}(\bro)+\rh_{\alpha}\bigl(\cle_{\az}(\bro)\bigr)
\ . \label{hhcq1}
\end{equation}
This inequality can be treated as an uncertainty relation
expressed in terms of the quantum $\alpha$-entropies of $\bro$ and
$\cle_{\az}(\bro)$. Certainty relations are formulated as upper
bounds on the sum of considered entropies. At this stage, only
simple bounds may be given. We merely recall that the quantum
$\alpha$-entropy is not more than the $\alpha$-entropy of the
completely mixed state. Thus, we obtain
\begin{equation}
H_{\alpha}(\az;\bro)+H_{\alpha}\bigl(\ax;\cle_{\az}(\bro)\bigr)\leq
2{\,}\rh_{\alpha}(\vro_{*})
\ , \label{hhcq2}
\end{equation}
where the completely mixed state $\vro_{*}=\pen/\Tr(\pen)$. We aim
to obtain uncertainty and certainty relations connected with the
purity $\Tr(\bro^{2})$ of the input state. In the following
sections, more detailed relations for two successive measurements
will be formulated in the qubit case.

In another scenario of successive measurements, the second
measurement is performed on the post-first-measurement state
conditioned on the actual measurement outcome. Let the first
measurement has given the outcome $z$. The probability of this
event is written as $p(z)=\Tr(\ppsf_{z}{\,}\bro)$. According to
the projection postulate, the state right after the measurement
is described by the projector $\ppsf_{z}$. In the second
measurement, therefore, the outcome $x$ is obtained with the
probability $p(x|z)=\Tr(\qqsf_{x}\ppsf_{z})$.\footnote{In the non-degenerate case.}
The latter is the
conditional probability of outcome $x$ given that the previous
measurement of $\az$ has resulted in $z$. In our case, the
function (\ref{hctm0}) is obviously expressed as
\begin{equation}
H_{\alpha}(\ax|z)=\frac{1}{1-\alpha}
\left(\sum_{{\,}x\in\spc(\ax)}\bigl(\Tr(\qqsf_{x}\ppsf_{z})\bigr)^{\alpha}
- 1\right)
\, . \label{haazb}
\end{equation}
It should be emphasized that this quantity does not depend on
$\bro$. For the scenario considered, an amount of uncertainties is
characterized by means of the conditional entropies
\begin{align}
H_{\alpha}(\ax|\az;\bro)&=\sum_{z\in\spc(\az)}
\bigl(\Tr(\ppsf_{z}{\,}\bro)\bigr)^{\alpha}{\,}H_{\alpha}(\ax|z)
\ , \label{ahct1}\\
\hw_{\alpha}(\ax|\az;\bro)&=\sum_{z\in\spc(\az)}
\Tr(\ppsf_{z}{\,}\bro){\,}H_{\alpha}(\ax|z)
\ . \label{ahct2}
\end{align}
Taking $\alpha=1$, both the $\alpha$-entropies are reduced to the
standard conditional entropy $H_{1}(\ax|\az;\bro)$. In Ref.
\cite{bfs2014}, the latter entropy was examined as a measure of
uncertainties in successive measurements. In the following, we
will give minimal and maximal values of the conditional entropies
(\ref{ahct1}) and (\ref{ahct2}) in the qubit case.

\section{First-scenario relations for successive qubit measurements}\label{sec4}

In this section, we will obtain uncertainty and certainty
relations for a qubit within the first scenario. The formulation
of uncertainty relations for successive measurements in terms of
the Shannon entropies was given in Ref. \cite{bfs2014}. The
authors of Ref. \cite{zzhang14} studied corresponding uncertainty
relations in terms of the R\'{e}nyi entropies. Although the
R\'{e}nyi and Tsallis entropies are closely connected, relations
for one of them are not immediately applicable to other. In
particular, the formula (\ref{exrt}) does not allow to move
between the R\'{e}nyi-entropy and Tsallis-entropy formulations of
the uncertainty principle. This fact also holds for the case of
successive measurements. We will formulate uncertainty relations
for two successive projective measurements in terms of Tsallis'
entropies.

Each density matrix in two dimensions can be represented in terms
of its Bloch vector as \cite{nielsen}
\begin{equation}
\bro=\frac{1}{2}{\,}\bigl(\pen+\vec{r}\cdot\vec{\bsg}\bigr)
\, . \label{bvrn}
\end{equation}
By $\vec{\bsg}$, we denote the vector of the Pauli matrices
$\bsg_{1}$, $\bsg_{2}$, $\bsg_{3}$. The three-dimensional vector
$\vec{r}=(r_{1},r_{2},r_{3})$ obeys $|\vec{r}|\leq1$, with
equality if and only if the state is pure. By calculations, the
obtain the purity
\begin{equation}
\Tr(\bro^{2})=\frac{1+|\vec{r}|^{2}}{2}
\ . \label{bvpy}
\end{equation}
The language of Bloch vectors is very convenient in description of
qubit states and their transformations \cite{nielsen,bengtsson}.
Following Ref. \cite{zzhang14}, we will aslo use this aproach in
representing projectors of the observables $\az$ and $\ax$.

Without loss of generality, we assume the observables to be
non-degenerate. Indeed, in two dimensions any degenerate
observable is inevitably proportional to the identity operator. We
will exclude this trivial case. Further simplification is reached
by rescaling eigenvalues of the observables. Working in
information-theoretic terms, we mainly deal with probability
distributions. In such a consideration, we can turn observables to
be dimensionless. Moreover, we can further shift eigenvalues of
the observables without altering the probabilities. Of course,
such actions are not appropriate for more traditional approach,
using the mean value and the variance. Thus, the spectral
decompositions are written as
\begin{align}
\az&=z_{+}\ppsf_{+}+z_{-}\ppsf_{-}
\ , \label{2dax}\\
\ax&=x_{+}\qqsf_{+}+x_{-}\qqsf_{-}
\ . \label{2day}
\end{align}
Taking the observables $\az$ and $\ax$ with eigenvalues $\pm1$, we
will arrive at dimensionless spin observables. Any projector
describes a pure state and, herewith, is represented by means of
some Bloch vector. We introduce two unit vectors $\vec{p}$ and
$\vec{q}$ such that
\begin{align}
\ppsf_{\pm}&=\frac{1}{2}{\,}\bigl(\pen\pm\vec{p}\cdot\vec{\bsg}\bigr)
\, , \label{pprn}\\
\qqsf_{\pm}&=\frac{1}{2}{\,}\bigl(\pen\pm\vec{q}\cdot\vec{\bsg}\bigr)
\, . \label{qqrn}
\end{align}
When eigenvalues of each of the observables are $\pm1$, we simply
have $\az=\vec{p}\cdot\vec{\bsg}$ and
$\ax=\vec{q}\cdot\vec{\bsg}$. In the first measurement, we measure
$\az$ in the pre-measurement state (\ref{bvrn}). The generated
probability distribution is written as
\begin{equation}
p(z=\pm1)=\frac{1\pm\vec{p}\cdot\vec{r}}{2}
\ . \label{fmpd}
\end{equation}
In the considered scenario, information contained in the qubit
after the first measurement is completely erased. By calculations,
we now have
\begin{align}
\cle_{\az}(\bro)&=\ppsf_{+}{\,}\bro{\,}\ppsf_{+}+\ppsf_{-}{\,}\bro{\,}\ppsf_{-}
\nonumber\\
&=\frac{1}{2}{\,}\bigl(\pen+(\vec{p}\cdot\vec{r}){\,}\vec{p}\cdot\vec{\bsg}\bigr)
\ , \label{clar}
\end{align}
with the Bloch vector $(\vec{p}\cdot\vec{r}){\,}\vec{p}$. Here,
the map turns the Bloch vector into its projection on $\vec{p}$.
The density matrix (\ref{clar}) describes the pre-measurement
state of the measurement of $\ax$. Similarly to Eq. (\ref{fmpd}),
we write generated probabilities in the form
\begin{equation}
p(x=\pm1)=\frac{1\pm(\vec{q}\cdot\vec{p})(\vec{p}\cdot\vec{r})}{2}
\ . \label{smpd}
\end{equation}
To avoid bulky expressions, we put the function of positive
variable with the parameter $\alpha>0$,
\begin{equation}
\eta_{\alpha}(\xi):=\frac{\xi^{\alpha}-\xi}{1-\alpha}
=-\xi^{\alpha}\ln_{\alpha}(\xi)
\ , \label{etdf}
\end{equation}
including $\eta_{1}(\xi)=-\xi\ln\xi$. Due to the space isotropy,
we can choose the frame of references in such a way that
$\vec{p}=\vec{e}_{3}$. Then the
entropic quantity (\ref{hhcq}) is written as
\begin{align}
&\sum_{m=\pm1}\eta_{\alpha}\!\left(\frac{1+m(\vec{p}\cdot\vec{r})}{2}\right)
+\sum_{n=\pm1}\eta_{\alpha}\!\left(\frac{1+n(\vec{q}\cdot\vec{p})(\vec{p}\cdot\vec{r})}{2}\right)
\nonumber\\
&=\sum_{m=\pm1}\eta_{\alpha}\!\left(\frac{1+m{\,}r_{3}}{2}\right)
+\sum_{n=\pm1}\eta_{\alpha}\!\left(\frac{1+n{\,}\mu{\,}r_{3}}{2}\right)
\, . \label{fnrz}
\end{align}
For brevity, we denote $\mu=(\vec{q}\cdot\vec{p})\in[-1;+1]$. To
obtain purity-based uncertainty and certainty relations, we will
search the minimum and the maximum of (\ref{fnrz}) under the
restriction that the Bloch vector length $|\vec{r}|$ is fixed. Due
to Eq. (\ref{bvpy}), the purity of a quantum state then remains
unchanged. The following statement takes place.

\newtheorem{prop1}{Proposition}
\begin{prop1}\label{pan1}
Let the length $|\vec{r}|$ of the Bloch vector of $\bro$ be fixed.
For all $\alpha>0$, the sum of $\alpha$-entropies for successive
measurements of non-degenerate observables $\az$ and $\ax$ is
bounded from below as
\begin{align}
&H_{\alpha}(\az;\bro)+H_{\alpha}\bigl(\ax;\cle_{\az}(\bro)\bigr)
\nonumber\\
&\geq
\sum_{m=\pm1}\eta_{\alpha}\!\left(\frac{1+m{\,}|\vec{r}|}{2}\right)
+\sum_{n=\pm1}\eta_{\alpha}\!\left(\frac{1+n{\,}\mu{\,}|\vec{r}|}{2}\right)
\, . \label{minr}
\end{align}
The equality in Eq. (\ref{minr}) holds if and only if $\bro$
commutes with $\az$, i.e., $\bro{\,}\az=\az{\,}\bro$. For all
$\alpha>0$, the sum of $\alpha$-entropies is bounded from above as
\begin{equation}
H_{\alpha}(\az;\bro)+H_{\alpha}\bigl(\ax;\cle_{\az}(\bro)\bigr)\leq
2\ln_{\alpha}(2)
\ . \label{maxr}
\end{equation}
The equality in Eq. (\ref{maxr}) holds if and
only if $\Tr(\az\bro)=\Tr(\az)/2$.
\end{prop1}

{\bf Proof.} For the given $|\vec{r}|$, the component
$r_{3}=(\vec{p}\cdot\vec{r})$ obeys
$-|\vec{r}|\leq{r}_{3}\leq+|\vec{r}|$. Since the right-hand side
of Eq. (\ref{fnrz}) is an even function of $r_{3}$, we can
restrict our consideration to the interval
$r_{3}\in\bigl[0;|\vec{r}|\bigr]$. For all $\alpha>0$, the
function (\ref{etdf}) is concave. Thus, the right-hand side of Eq.
(\ref{fnrz}) is concave with respect to $r_{3}$. A concave
function reaches the minimal value at one or more least points of
the interval. To minimize the function (\ref{fnrz}), we should
therefore compare its values for $r_{3}=0$ and for
$r_{3}=|\vec{r}|$. The latter actually leads to the minimum,
whence the claim (\ref{minr}) is proved. Indeed, substituting
$r_{3}=0$ leads to the uniform distribution and, therefore, to the
maximal value $\ln_{\alpha}(2)$ of each of two entropies. The last
comment justifies the claim (\ref{maxr}).

We shall now prove conditions for the equality. Substituting
$r_{3}=\pm|\vec{r}|$, the right-hand side of Eq. (\ref{fnrz}) is
equal to the right-hand side of Eq. (\ref{minr}). Some inspection
shows that deviating $r_{3}$ from the points $\pm|\vec{r}|$ will
certainty increase the term (\ref{fnrz}). Thus, these points are
the only case when the inequality (\ref{minr}) is saturated. In
this case, we have $\vec{r}\parallel\vec{p}$, whence the density
matrix is diagonal with respect to the common eigenbasis of the
projectors $\ppsf_{\pm}$. Hence, the operators $\bro$ and $\az$
commute. To saturate the inequality (\ref{maxr}), the two
entropies must reach the maximal value. The only case is
$r_{3}=0$, when the probability distributions $(1\pm{r}_{3})/2$
and $(1\pm\mu{\,}r_{3})/2$ are both uniform. Since
\begin{equation}
\Tr\bigl(
(\vec{p}\cdot\vec{\bsg})(\vec{r}\cdot\vec{\bsg})
\bigr)=2(\vec{p}\cdot\vec{r})
\ , \label{prbb}
\end{equation}
for $\vec{r}\perp\vec{p}$ we obtain $\Tr(\ppsf_{\pm}\bro)=1/2$.
Combining the latter with (\ref{2dax}) gives
$\Tr(\az\bro)=\Tr(\az)/2$. $\square$

The result (\ref{minr}) is a Tsallis-entropy family of uncertainty
relations for successive projective measurements of a qubit. The
pre-measurement density matrix $\bro$ describes a mixed state of
the purity (\ref{bvpy}). For the given $|\vec{r}|$, this purity is
constant. In two dimensions, a geometrical description in terms of
the Bloch vector is more convenient. The equality in (\ref{minr})
takes place, if and only if $\vec{r}\parallel\vec{p}$. Among
states of the fixed purity, minimal uncertainties are revealed by
states, whose Bloch vector is collinear to the Bloch vector
associated with the projectors on the eigenspaces of $\az$. In
terms of operators, this condition implies the commutativity of
$\az$ and $\bro$. Further, the density matrix is then a fixed
point of $\cle_{\az}$, i.e., $\cle_{\az}(\bro)=\bro$ for
$\vec{r}\parallel\vec{p}$.

We further note that the lower bound (\ref{hhcq1}) is typically
not saturated for $\vec{r}\parallel\vec{p}$. Indeed, the first
entropic sum in the quantity (\ref{minr}) actually becomes equal
to the quantum $\alpha$-entropy of $\bro$. This is not the case
for the second one. In general, the second entropic sum in the
formula (\ref{minr}) is  strictly larger than the quantum
$\alpha$-entropy of $\cle_{\az}(\bro)$. The equality takes place
only for $\mu=\pm1$ that is equivalent to
$\vec{q}\parallel\vec{p}$. In such a situation, the operators
$\bro$, $\az$, and $\ax$ are all diagonal in the same common
eigenbasis. Then the picture becomes purely classical in
character.

As we see, uncertainties in considered successive measurements are
minimized for $\vec{r}\parallel\vec{p}$. It is natural to expect
that the condition $\vec{r}\perp\vec{p}$ will lead to an opposite
case of maximal uncertainties. In effect, this case actually lead
to the maximal values to both the entropies of the left-hand side
of (\ref{maxr}). Then the quantum operation $\cle_{\az}$ maps the
input $\bro$ into the completely mixed state. Here, we have
$\cle_{\ax}\circ\cle_{\az}(\bro)=\cle_{\az}(\bro)=\vro_{*}$. Thus,
the upper bound (\ref{hhcq2}) is actually saturated. When the two
eigenvalues of $\az$ are symmetric with respect to $0$, as for
spin observables, the condition $\vec{r}\perp\vec{p}$ implies that
the mean value of $\az$ in the state $\bro$ is zero, i.e.,
$\Tr(\az\bro)=0$. Indeed, in such a case we deal with the
traceless observable.

\section{Bounds for the second scenario of successive qubit measurements}\label{sec5}

In this section, we will obtain uncertainty and certainty
relations for a qubit within the second scenario. Here, the second
measurement is performed on the post-first-measurement state
conditioned on the actual measurement outcome. Using the
conditional R\'{e}nyi entropy, corresponding uncertainty relations
were derived in Ref. \cite{zzhang14}. To quantify uncertainties,
we will use the conditional Tsallis entropies (\ref{hct1}) and
(\ref{hct2}). They cannot be related immediately to the
conditional R\'{e}nyi entropy. So, a formulation in terms the
conditional entropies (\ref{hct1}) and (\ref{hct2}) is of own
interest. It turns out that the entropy (\ref{hct1}) rather gives
a more sensitive measure than (\ref{hct2}). Entropic certainty
bounds for successive measurements seem to be not studied in the
literature.

We first measure the observable $\az$ in the state (\ref{bvrn}),
for which the probabilities of outcomes is calculated according to
Eq. (\ref{fmpd}). If the first measurement has given the outcome
$m$, then the post-first-measurement state is described by the
projector $\ppsf_{m}$ with the Bloch vector $m\vec{p}$. Then we
perform the measurement of $\ax$ with generating the probability
distribution
\begin{equation}
p(x=n|z=m)=\frac{1+n(\vec{q}\cdot{m}\vec{p})}{2}
\ , \label{gepd}
\end{equation}
where $m,n=\pm1$. These quantities are conditional probabilities
used in Eq. (\ref{haazb}). For both $m=\pm1$, the entropic
function (\ref{haazb}) is expressed as
\begin{equation}
H_{\alpha}(\ax|z=m)=\rh_{\alpha}\bigl(\cle_{\ax}(\ppsf_{m})\bigr)
=\sum_{n=\pm1}\eta_{\alpha}\!\left(\frac{1+n{\,}\mu}{2}\right)
\, . \label{gehf}
\end{equation}
Hence, the two conditional Tsallis entropies are represented as
\begin{align}
H_{\alpha}(\ax|\az;\bro)&=
\sum_{m=\pm1}\left(\frac{1+m{\,}r_{3}}{2}\right)^{\!\alpha}
\sum_{n=\pm1}\eta_{\alpha}\!\left(\frac{1+n{\,}\mu}{2}\right)
\, , \label{fcen}\\
\hw_{\alpha}(\ax|\az;\bro)&=
\sum_{n=\pm1}\eta_{\alpha}\!\left(\frac{1+n{\,}\mu}{2}\right)
\, . \label{scen}
\end{align}
Thus, the second form of conditional $\alpha$-entropies does not
depend on the input state $\bro$. This entropic quantity is
completely given by taking the observables $\az$ and $\ax$. Except
for $\alpha=1$, the first conditional entropy (\ref{fcen}) depends
on the state $\bro$ and both the observables. The right-hand side of Eq.
(\ref{fcen}) gives a general expression of the entropy. Let us examine an
interval, in which this quantity ranges as a function of inputs of
the fixed purity.

\newtheorem{prop2}[prop1]{Proposition}
\begin{prop2}\label{pan2}
Let the length $|\vec{r}|$ of the Bloch vector of $\bro$ be fixed.
For $\alpha\in(0;1)$, the conditional $\alpha$-entropy for
successive measurements of non-degenerate observables $\az$ and
$\ax$ obeys
\begin{align}
&\bigl(1+(1-\alpha)\rh_{\alpha}(\bro)\bigr)\sum_{n=\pm1}\eta_{\alpha}\!\left(\frac{1+n{\,}\mu}{2}\right)
\leq{H}_{\alpha}(\ax|\az;\bro)
\ , \label{cea011}\\
&H_{\alpha}(\ax|\az;\bro)\leq2^{1-\alpha}\sum_{n=\pm1}\eta_{\alpha}\!\left(\frac{1+n{\,}\mu}{2}\right)
\, . \label{cea012}
\end{align}
The lower bound (\ref{cea011}) is reached if and only if $\bro$
commutes with $\az$, i.e., $\bro{\,}\az=\az{\,}\bro$. The upper
bound (\ref{cea012}) is reached if and only if
$\Tr(\az\bro)=\Tr(\az)/2$.

For $\alpha\in(1;\infty)$, the conditional $\alpha$-entropy for
successive measurements of non-degenerate observables $\az$ and
$\ax$ obeys
\begin{align}
&2^{1-\alpha}\sum_{n=\pm1}\eta_{\alpha}\!\left(\frac{1+n{\,}\mu}{2}\right)\leq{H}_{\alpha}(\ax|\az;\bro)
\ , \label{cea021}\\
&H_{\alpha}(\ax|\az;\bro)
\leq\bigl(1+(1-\alpha)\rh_{\alpha}(\bro)\bigr)\sum_{n=\pm1}\eta_{\alpha}\!\left(\frac{1+n{\,}\mu}{2}\right)
\, . \label{cea022}
\end{align}
The lower bound (\ref{cea021}) is reached if and only
$\Tr(\az\bro)=\Tr(\az)/2$. The upper bound (\ref{cea022}) is
reached if and only if $\bro{\,}\az=\az{\,}\bro$.
\end{prop2}

{\bf Proof.} We need only find those least values that exactly
bound the first sum in the right-hand side of Eq. (\ref{fcen}).
For brevity, we define the function
\begin{equation}
g_{\alpha}(r_{3}):=\left(\frac{1+r_{3}}{2}\right)^{\!\alpha}+\left(\frac{1-r_{3}}{2}\right)^{\!\alpha}
\, . \label{dfga}
\end{equation}
As this function is even, the aim is to find its minimum and
maximum on the interval $r_{3}\in\bigl[0;|\vec{r}|\bigr]$.

For $\alpha\in(0;1)$, the function
$r_{3}\mapsto{g}_{\alpha}(r_{3})$ is concave. So, we have
$g_{\alpha}(r_{3})\leq2^{1-\alpha}=g_{\alpha}(0)$ by using
Jensen's inequality with the weights $1/2$. Hence, the claim
(\ref{cea012}) follows. Due to concavity, the minimum is reached
for one of two least points of the interval
$r_{3}\in\bigl[0;|\vec{r}|\bigr]$. So, the desired minimum is
\begin{align}
g_{\alpha}\bigl(|\vec{r}|\bigr)&=
\left(\frac{1+|\vec{r}|}{2}\right)^{\!\alpha}+\left(\frac{1-|\vec{r}|}{2}\right)^{\!\alpha}
\nonumber\\
&=1+(1-\alpha)\rh_{\alpha}(\bro)
\ , \label{gadf}
\end{align}
as the eigenvalues of $\bro$ are $\bigl(1\pm|\vec{r}|\bigr)/2$.
Combining this with Eq. (\ref{fcen}) finally gives Eq.
(\ref{cea011}). Focusing on the conditions for equality,
$r_{3}=\pm|\vec{r}|$ is equivalent to $\vec{r}\parallel\vec{p}$,
and $r_{3}=0$ is equivalent to $\vec{r}\perp\vec{p}$. The former
implies $\bro{\,}\az=\az{\,}\bro$, the latter implies
$\Tr(\az\bro)=\Tr(\az)/2$.

For $\alpha\in(1;\infty)$, the function
$r_{3}\mapsto{g}_{\alpha}(r_{3})$ is convex. Now, we have
$g_{\alpha}(r_{3})\geq2^{1-\alpha}=g_{\alpha}(0)$ by using
Jensen's inequality with the weights $1/2$. By convexity, the
maximum is reached for one of two least points of the interval
$r_{3}\in\bigl[0;|\vec{r}|\bigr]$. In this case, the right-hand
side of Eq. (\ref{gadf}) represents the desired maximum. The
conditions for equality are treated similarly. $\square$

The formulas (\ref{cea011}) and (\ref{cea021}) describe the
minimal values of the conditional $\alpha$-entropy (\ref{fcen})
for the states of fixed purity. The bounds (\ref{cea012}) and
(\ref{cea022}) present the maximal values. Unlike Eq. (\ref{fcen}), 
the inequalities (\ref{cea011})--(\ref{cea022}) do not depend on 
a direction of the Bloch vector. For the given
observables $\az$ and $\ax$, the mentioned results involve the
constant factor equal to Eq. (\ref{gehf}). For the case
$\alpha\in(0;1)$, the conditional entropy is minimized or
maximized under the same conditions as the sum of two
$\alpha$-entropies in the first scenario (see Proposition
\ref{pan1} above). For $\alpha\in(1;\infty)$, the conditions for
the equality are swapped. For $\alpha=1$, both the forms of
conditional entropies give
\begin{equation}
H_{1}(\ax|\az;\bro)=\sum_{n=\pm1}\eta_{1}\!\left(\frac{1+n{\,}\mu}{2}\right)
\leq\ln2
\, . \label{fcen1}
\end{equation}
Using the standard conditional entropy, this formula expresses an
upper bound for the second scenario of two successive
measurements. Overall, we can say the following. The conditional
$\alpha$-entropy (\ref{fcen}) seems to be more appropriate, since
it depends also on the pre-measurement state. This state is not
involved by the conditional $\alpha$-entropy (\ref{scen})
including the standard case (\ref{fcen1}). However, such a
property is two-dimensional in character. In more dimensions,
entropic functions of the form (\ref{haazb}) will generally depend
on the label $z$. Hence, the $\alpha$-entropy (\ref{ahct2}) will
be dependent on the pre-measurement state.

We shall now consider the entropic quantity (\ref{gehf}). It
reaches its maximal value $\ln_{\alpha}(2)$ for
$\vec{p}\perp\vec{q}$, when $\mu=0$. Here, we have arrived at an
interesting observation. For the given input $\bro$, both the
measures (\ref{fcen}) and (\ref{scen}) are maximized, when the
eigenbases of $\az$ and $\ax$ are mutually unbiased. The condition
$\vec{p}\perp\vec{q}$ actually implies that
$\bigl|\langle{z}|x\rangle\bigr|=1/\sqrt{2}$ for all labels $z$
and $x$. For instance, this property takes place for eigenbases of
any two of the three Pauli matrices. Such eigenstates are
indistinguishable in the following sense. The detection of a
particular basis state reveals no information about the state,
which was prepared in another basis. Indeed, two possible outcomes
are then equiprobable. As is known, this property is used in the
BB84 protocol of quantum key distribution \cite{bb84}.

Thus, formulation of certainty relations for successive
measurement again emphasizes a role of mutual unbiasedness. The
concept of mutually unbiased bases is naturally posed in
arbitrary finite dimensions. If two bases are mutually unbiased,
then the overlaps between any basis state in one basis and all
basis states in the other are the same. Mutually unbiased bases
have found use in many questions of quantum information. They also
connected with important mathematical problems (see the review
\cite{bz10} and references therein). Some of the above conclusions
can be extended to an arbitrary dimensionality. The
generalization is formulated as follows.

\newtheorem{prop3}[prop1]{Proposition}
\begin{prop3}\label{pan3}
Let $\az$ and $\ax$ be two non-degenerate $d$-dimensional
observables. For each density $d\times{d}$-matrix $\bro$, the
conditional entropies of successive measurements of $\az$ and
$\ax$ obey
\begin{align}
H_{\alpha}(\ax|\az;\bro)
&\leq\Tr\bigl(\cle_{\az}(\bro)^{\alpha}\bigr){\,}\ln_{\alpha}(d)
\ , \label{pr31}\\
\hw_{\alpha}(\ax|\az;\bro)&\leq\ln_{\alpha}(d)
\ . \label{pr32}
\end{align}
If the eigenbases of $\az$ and $\ax$ are mutually unbiased, 
then the equality is reached in both Eqs. (\ref{pr31}) and (\ref{pr32}). 
For strictly positive $\bro$, mutual unbiasedness of 
the eigenbases is the necessary condition for equality.
\end{prop3}

{\bf Proof.} By $\bigl\{|z\rangle\}$ and $\bigl\{|x\rangle\}$, we
respectively denote eigenbases of the non-degenerate observables
$\az$ and $\ax$. Recall that the Tsallis $\alpha$-entropy is not
larger than $\ln_{\alpha}(d)$, where $d$ is the number of
outcomes. For each eigenvalue $z$, we then have
\begin{equation}
H_{\alpha}(\ax|z)\leq\ln_{\alpha}(d)
\ . \label{hxzml}
\end{equation}
In the case considered, we clearly have
$\sum_{z}p(z)^{\alpha}=\Tr\bigl(\cle_{\az}(\bro)^{\alpha}\bigr)$.
Combining the latter with Eq. (\ref{hxzml}) provides the claims
(\ref{pr31}) and (\ref{pr32}). We shall now proceed to the
conditions for equality. The maximal entropic value
$\ln_{\alpha}(d)$ is reached only for the uniform distribution,
when the probabilities are all $1/d$. To saturate Eq.
(\ref{hxzml}) with the given $z$, we should therefore have
\begin{equation}
\forall{\,}x\in\spc(\ax):
{\>}\bigl|\langle{x}|z\rangle\bigr|=\frac{1}{\sqrt{d}}
\ . \label{xtwb}
\end{equation}
When $\bro$ is strictly positive, $p(z)\neq0$ for all
$z\in\spc(\az)$. To reach the equality in both Eqs. (\ref{pr31})
and (\ref{pr32}), the condition (\ref{xtwb}) should be provided
for all $z\in\spc(\az)$. Then the two eigenbases are mutually
unbiased. $\square$

Although the BB84 scheme of quantum cryptography is primarily
important, other protocols have been studied in the literature
\cite{grtz02}. Some of them are based on mutually unbiased bases
(see, e.g., Refs. \cite{bruss98,mdg13}). We have seen above that
the conditional entropies for a pair of successive projective
measurements are maximized just in the case of mutual
unbiasedness. Entropic uncertainty relations for several mutually
unbiased bases were studied in Refs.
\cite{ballester,molm09,MWB10,rast13b}. It would be interesting to
consider main questions of this section for many mutually unbiased
bases. Such investigations may give an additional perspective of
possible use of mutually unbiased bases in quantum information
processing. It could be a theme of separate investigation.

\section{Conclusion}\label{sec6}

We have studied Tsallis-entropy uncertainty and certainty
relations for two subsequent measurements of a qubit. Despite of
very wide prevalence of the Heisenberg principle, there is no
general consensus over its scope and validity. The following claim
is commonly accepted. It is not possible to assign jointly exact
values for two or more incompatible observables. There are several
ways to fit this claim as a quantitative statement. Heisenberg's
original argument is adequately formulated in terms of noise and
disturbance \cite{ozawa04}. Using the two scenarios of successive
measurements, we are able to fit the question of measuring
uncertainties in a different way. Such an approach may be more
significant in the sense of its potential applications in studying
protocols of quantum information. Indeed, subsequent manipulations
with qubits rather deal with an output state of the latter stage.
Uncertainty relations for two successive measurements were already
examined in terms of the standard entropies \cite{bfs2014} and the
R\'{e}nyi entropies \cite{zzhang14}. At the same time, certainty
relations for successive measurements seem to be not addressed in
the literature.

The following two scenarios of successive measurements were
considered. In the first scenario, a subsequent measurement is
performed on the quantum state generated after the previous stage
with completely erased information. In the second scenario, a
subsequent measurement is performed on the post-measurement state
conditioned on the actual measurement outcome. for both the
scenarios, we derived uncertainty and certainty bounds on
$\alpha$-entropic functions related to successive measurements of
a pair of qubit observables. The conditions for equality in these
bounds were obtained as well. Some of found results in the second
scenario were extended to arbitrary finite dimensionality. They
are connected with a frequently used property of mutually unbiased
bases. In effect, the detection of a particular basis state
reveals no information about the state, which was prepared in
another basis. It would be interesting to study uncertainty and
certainty relations for successive measurements in  several
mutually unbiased bases.

\end{document}